\documentclass[a4paper,twocolumn,fleqn,usenatbib]{mnras}

\usepackage{bm}
\usepackage{amsmath}
\usepackage[T1]{fontenc}
\usepackage{ae,aecompl}
\usepackage{adjustbox}
\usepackage{graphicx}	
\usepackage{amssymb}	
\usepackage{color}
\usepackage{float}
\usepackage{txfonts}
\usepackage{units}
\usepackage{url}
\usepackage{slantsc}

\def\mhcm{\rm m_{H}~cm^{-3}}
\def\msun{M_{\odot}}
\def\msunyr{{\rm M_{\odot} yr^{-1}}}
\def\cmc{{\rm cm^{-3}}}
\def\vw{v_{\rm out}}  
\def\vdn{v_{\rm DN}}
\def\eps{\epsilon_{\rm s}}
\def\epe{\epsilon_{\rm e}}
\def\epeo{\bar{\epsilon}_{\rm e,-1}}
\def\epb{\epsilon_{\rm B}}
\def\epbt{\epsilon_{\rm B,-2}}
\def\mcnm{M_{\rm CNM}}
\def\gmm{\gamma_m}
\def\rhow{\rho_{\rm out}}
\def\cv{c_{f}}
\def\Csgr{C_{\rm sgr}}
\def\etat{\eta_{-3}}
\def\ddp{{\rm d}p}
\def\Lkin{L_{\rm kin}}
\def\Mw{\dot{M}_{\rm out}}
\def\Rc{R_{\rm c}} 
\def\ergs{{\rm erg~s^{-1}}}
\def\ergshz{{\rm erg~s^{-1}Hz^{-1}}}
\def\kbow{k_{\rm bow}}
\def\tbst{t_{\rm bst}}
\def\Gm{\Gamma^{'}}
\def\nutp{\nu_{\rm tp}}
\def\be{\begin{equation}}
\def\ee{\end{equation}}

\title[Radio Emission]{Radio Emission from Outflow-Cloud Interaction and Its Constraint on TDE Outflow}

\author[Mou et~al.]{Guobin~Mou,$^{1,2}$\thanks{gbmou@whu.edu.cn (GM)}
 Tinggui Wang,$^{3}$\thanks{twang@ustc.edu.cn (TW)}
 Wei Wang,$^{1,2}$\thanks{wangwei2017@whu.edu.cn (WW)}
 Jingjing Yang$^{1}$
\\
$^{1}$School of Physics and Technology, Wuhan University, Wuhan 430072, China  \\
$^{2}$WHU-NAOC Joint Center for Astronomy, Wuhan University, Wuhan 430072, China \\
$^{3}$School of Astronomy and Space Science, University of Science and Technology of China, Hefei 230026, China \\
}

\date{in original form 2021 Aug 30}

\pubyear{ -- }

\begin{document}
\label{firstpage}
\pagerange{\pageref{firstpage}--\pageref{lastpage}}
\maketitle

\begin{abstract} 
Tidal disruption event (TDE) can launch an ultrafast outflow. 
If the black hole is surrounded by large amounts of clouds, outflow-cloud interaction will generate bow shocks, accelerate electrons and produce radio emission. 
Here we investigate the interaction between a non-relativistic outflow and clouds in active galaxies, which is manifested as outflow-BLR  (broad line region) interaction, and can be extended to outflow-torus interaction. 
This process can generate considerable radio emission, which may account for the radio flares appearing a few months later after TDE outbursts. Benefitting from efficient energy conversion from outflow to shocks and the strong magnetic field, outflow-cloud interaction may play a non-negligible, or even dominating role in generating radio flares in a cloudy circumnuclear environment if the CNM density is no more than 100 times the Sgr A*-like one.  
In this case, the evolution of radio spectra can be used to directly constrain the properties of outflows.   
\end{abstract} 
 

\section{Introduction}
Apart from releasing strong and transient electromagnetic radiations, TDEs also produce fast and energetic outflow in two mechanisms: the violent self-interaction due to general relativistic apsidal precession (\citealt{sadowski2016, lu2020}), or the super-Eddington accretion phase (\citealt{dai2018,curd2019}).  
Observationally, the existence of TDE outflow can be confirmed directly by UV and X-ray spectra (\citealt{yang2017,gezari2021}), or inferred indirectly by radio emission (e.g., \citealt{alexander2020}). 

To date, about eleven of TDE candidates exhibit radio emission of $10^{36-42}~\ergs$ with time lags spanning from days to years (\citealt{alexander2020}). 
The cosmic ray electrons (CRe) accounting for the radio emission are thought to be accelerated in forward/external shock driven by TDE outflow interacting with circumnuclear medium (CNM), or shock driven by jets (\citealt{bloom2011, burrows2011, zauderer2011}). 
\citet{giannios2011} explored the observational consequences of jet-CNM interactions, and predicted that bright radio emission can be produced in jetted TDE and may be detectable for years, which also contains the information on CNM density.     
For outflow-CNM model, \citet{duran2013} provided a method for estimating the minimal energy of CRe and magnetic field components for the observed synchrotron emission, which should be regarded as a strict lower limit for shock energy and outflow kinetic energy. 
When applying this model to the known radio TDEs (e.g., \citealt{matsumoto2021}), it is found that those radio flares favor a CNM with density much higher than that of Sgr A* in the innermost region of $\lesssim 10^{17}$cm.  
However, the hot diffuse medium may not be the only component of CNM. In active galactic nucleus (AGN), normally there exists a so-called broad line region (BLR) composed of large amounts of clouds surrounding the central black hole. 
For a TDE occurring in a SMBH surrounded by clouds, apart from outflow-CNM interaction, the outflow-cloud interaction will yield bow shocks, and convert the outflow kinetic energy into shock energy efficiently. 
This process also accelerates relativistic electrons, and generates synchrotron radiation which may dominate the overall radio emission in many cases (as this work will prove). 
Interestingly, as pointed in \citet{alexander2020}, most of the TDEs with detected low-luminosity radio emission occurring in galaxies with signs of recent or ongoing AGN activity. 
Moreover, the interaction of falling back debris with pre-existing AGN may also be directly manifested in TDE light curves, e.g., dimming the associated flare (debris-corona interaction, \citealt{bonnerot2016}), or causing a sudden drop in the light curve (debris-disk interaction, \citealt{kath2017}). 
Our study is aimed to investigate the outflow-cloud interaction. Hereafter, we refer to CNM specifically as the hot diffuse medium, to distinguish it from the condensed clouds. We do not use the term ``TDE wind'' as in our previous articles (referring to the large opening-angle outflow, \citealt{mou2021,mou2021b}), since the current results also apply to collimated jet as long as it's non-relativistic. 

In Section 2, we introduce outflow-BLR interaction and synchrotron radiation. Section 3 shows model applications, in which we simplify the formula, present radio constraint on outflow, and apply it to three TDEs. We discuss outflow-cloud and outflow-CNM models in Section 4, and give a brief summary in Section 5. Notation $Q_x$ is equivalent to $Q/10^x$ in cgs units unless otherwise specified (exceptions, $t_2\equiv t/10^2$days, $\Omega_4 \equiv \Omega/4$sr and $r_{-2}\equiv r/10^{-2}$pc).

\section{TDE Outflow and Cloud interaction}

\subsection{Broad Line Region}  
The BLR is a fundamental component in AGN, which is responsible for the broad emission lines with typical FWHM of several thousand km~s$^{-1}$ in AGN spectra. 
The origin of the BLR is still an open question.  
Observations suggest that the broad lines disappear when the AGN luminosity drops below $5\times 10^{39}~\ergs (M_{\rm BH}/10^7\msun)^{2/3}$, while some ``true'' Seyfert 2 galaxies with higher Eddington ratios do exist (\citealt{elitzur2009}).   
Statistics of nearby galaxies indicate that SMBHs with luminosities above this value are widespread (over half of the sources satisfying the condition, see, e.g., \citealt{ho2009}).  
Although originally inferred from the AGN spectra, the presence of BLRs in quiescent galaxies is largely unknown. 
There are indeed several cloud-like objects (unclear nature) lying within 0.04 pc of Sgr A* (\citealt{ciurlo2020}), including the G2 cloud plunging into the innermost region of $10^3$ Schwarzschild radius (\citealt{gillessen2012}).  
Here our outflow-cloud interaction model is manifested as outflow-BLR interaction, but we should remind that it can also be applied to ``hidden'' clouds around quiescent SMBHs.  

\begin{figure}
\includegraphics[width=0.95\columnwidth]{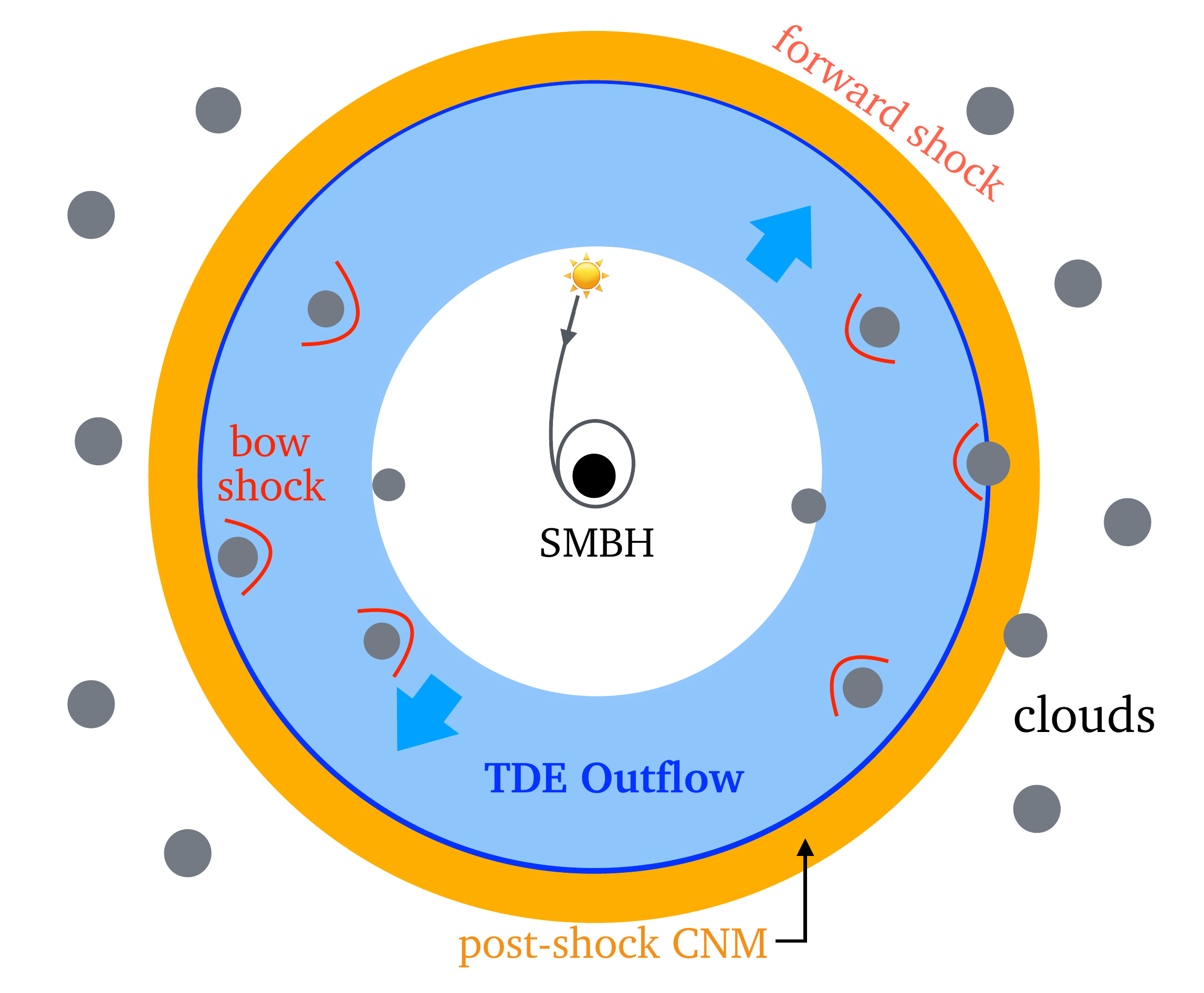}
 \caption{Sketch of the interaction between the TDE outflow and BLR. The light blue ``shell'' marks the TDE outflow (after outflow's launch), and the grey circles mark the BLR clouds. The red arcs represent the bow shocks. The orange shell represents the post-shock CNM. Our model concerns the synchrotron emission of CRe accelerated by bow shocks.  } 
 \label{fig1}
\end{figure}

The BLR is regarded to an ensemble of small and optically thick clouds with density $\sim 10^{10}~\mhcm$. 
Reverberation mapping measurements of Balmer lines suggest that the mean radii are $10^{0}-10^{1}$ light-days for bolometric luminosities of $10^{43}-10^{45}~\ergs$ (\citealt{kaspi2005}). The distribution of the clouds may favor a thick disk geometry with a large opening angle (e.g., \citealt{gravity2018}). 
The covering factor of BLR is in the order of 0.1 (\citealt{netzer2013}).  
However, when dealing with the interaction between the transient TDE outflow and clouds, we should use the \emph{effective} covering factor $\cv$, of which only part of the clouds are simultaneously interacting with the outflow or outflow ``shell'' (after outflow's launch, see figure \ref{fig1}). The $\cv$ should be less than the global one, and it increases over time when the outflow begins to encounter the BLR. 
The duration of launching energetic outflow $\tbst$ is in the order of 1 month, since neither the violent self-interaction nor super-Eddington accretion is able to sustain too long (see \citealt{mou2021}). After $\tbst$, the outflow probably still emanate, but its power may be significantly reduced.  

\subsection{Comic Ray Electrons}

The fraction of outflow kinetic energy converted into CRe depends on two factors:
the fraction of outflow's energy converted into shock -- $\eps$, and the fraction of shock energy converted into CRe -- $\epe$.
For BLR clouds, $\eps$ is approximately equal to $\cv$. 

According to the studies on SNRs (e.g., \citealt{blasi2013}) and numerical simulations (\citealt{caprioli2014}), about $10\%$ of the shock energy can be converted into cosmic rays. 
CRe follows a power-law form of Lorentz factor (LF) as $dN(\gamma_e)/d\gamma_e = A \gamma^{-p}_e$ ($p>2.0$), and the index $p$ is typically 2.5--3.0 for Newtonian shocks (e.g., \citealt{chevalier2006}), or 2.1--2.5 in mildly relativistic shocks. 
For a bulk motion with LF of $\gamma$, the minimal LF of CRe $\gmm$ is derived by: $\gmm = \epe \frac{m_p}{m_e}\frac{p-2}{p-1}(\gamma-1)$ (e.g., \citealt{duran2013}). 
Most of the CRe energy is contributed by those around $\gmm$. 
For the so-called ``deep-Newtonian'' phase (non-relativistic bulk motion of $v \leqslant \vdn$, \citealt{huang2003}), however, $\gmm$ will reduce to $<2$, and most of the shock-accelerated electrons are non-relativistic. 
The critical bulk velocity is given by 
$\vdn=c \sqrt{16 m_e/(m_p\bar{\epsilon}_e)}$ from solving $\gmm=2$, where $\bar{\epsilon}_e \equiv 4\epe(p-2)/(p-1)$. 
If $\bar{\epsilon}_e=0.1$, we have $\vdn=0.3c$.
In deep-Newtonian regime, the electron energy distribution peaks at $\gamma_e \sim 2$, as a result of CRe following a power-law distribution in momentum (instead of energy) with slope $p$ (\citealt{sironi2013}), and only a fraction $(v/\vdn)^2$ of electrons in post-shock gas participate in the power-law distribution with LF $\gamma_e \gtrsim 2$ and emit synchrotron radiation. 
When $v > \vdn$, $\gmm \simeq 230 \bar{\epsilon}_e (v/c)^2 > 2$. Similar to \citet{matsumoto2021}, we treat the two regimes with $\vw \leqslant \vdn$ and $\vw > \vdn$ separately. \footnote{Our $v_{\rm DN}$ is slightly different from \citet{matsumoto2021} in which $\vdn=c \sqrt{8 m_e/(m_p \bar{\epsilon}_e)}$ . } 

For outflow-cloud interaction, the high pressure behind the bow shock drives the shocked outflow to expand towards the back of the cloud, in which its thermal energy is converted into its kinetic energy.
Thus, the CRs transported with the fluid suffers from the adiabatic loss. 
In \citet{mou2021b}, we proved that behind the bow shock, the adiabatic cooling timescale of CR due to transport with fluid (ignoring diffusion) is $t_{\rm ad}\simeq \kbow R_c/\vw \sim 10^5{\rm s}~ R_{c,13} v_9^{-1}$, where $R_c$ is the size of the cloud, and $\kbow \simeq 10$. 
\footnote{We draw attention to the slight difference. In \citet{mou2021b}, $R_c$ is the cloud radius (half the size), and $\kbow=20$ for outflow duration $\gg R_c/\vw$ (see appendix B therein). This is equivalent to the expression of $t_{ad}$ here. } 
If the duration of a bow shock is longer than $t_{\rm ad}$ (which almost always stands in our concerns here), the CR energy of this shock will saturate, and remains at the value of the injection rate times $t_{\rm ad}$. 
In this context, the total number $N_e$ of the electrons with LF $\gamma \gtrsim \gmm$,  
can be estimated by 
\be
E_{\rm cre}= N_e m_e c^2 \gmm \frac{p-1}{p-2} \simeq \cv \epe \Lkin t_{\rm ad}
\label{ne}
\ee
where $\Lkin=\Omega r^2 \rhow(r) \vw^3/2$ is the kinetic luminosity of TDE outflow, $\Omega$ is the solid angle of the outflow, $\rhow$ and $\vw$ are the outflow density and velocity, respectively, and $\gmm=2$ for $\vw \leqslant \vdn$, or $23 \epeo \beta^2$ for $\vw > \vdn$ ($\beta \equiv \vw/c$). 
These electrons originate from outflow swept by bow shocks, instead of CNM swept by forward shock. 
The expression of the coefficient $A$ in $dN(\gamma_e)/d\gamma_e$ (in unit volume) is derived by:
\be
A=(p-1) \gmm^{p-1} \frac{\rhow(r)}{m_p} \cdot \Lambda
\label{Acoeff}
\ee
where $\Lambda \equiv \min[(\vw/\vdn)^2, 1]$.

\subsection{Magnetic Field}
Due to the magnetic field amplification (\citealt{bell2001, schure2012}), the magnetic pressure is enhanced to $\sim 10^{-2} \rhow \vw^2$ in shock downstream (\citealt{volk2005}). 
Setting $B^2/8\pi = \epb \rhow \vw^2$ and adopting $r=\vw t$, we have 
\be 
B = 14.6 ~{\rm Gauss} ~\epbt^{\frac{1}{2}} \dot{m}^{\frac{1}{2}} \Omega^{-\frac{1}{2}} v_9^{-\frac{1}{2}} t_2^{-1}  ~,
\label{bmg}
\ee
where $\epb=0.01$ is the fiducial parameter, $t_2\equiv t/10^2$days, and $\dot{m}\equiv \Mw/1\msunyr$, $\Mw=\Omega r^2 \rhow \vw$ is the mass outflow rate of TDE outflow. 
When expressing $B$ in the form of time by virtue of $r=\vw t$, it means that we simplified the outflow-cloud interaction as a process occurring at the same radius. This is valid when $t$ exceeds $\tbst$ by several times or more and the transient outflow forms a thin ``shell''. 
Before that, although outflow-cloud interaction for the outflow's tail end also generates radio emission, the bow shock energy of the strongly compressed cloud drops greatly and becomes negligible. The compression timescale is $R_c/(\chi^{-0.5}\vw)=12{\rm d}~ R_{c,13} v_9^{-1} \chi_4^{-1/2}$ ($\chi$ is the density ratio of cloud to outflow, \citealt{mckee1975}), which is typically shorter than the age of radio flares $t$.  
However, if the cloud compression timescale is indeed comparable to $t$, the contribution of outflow's tail would not be neglected, which will affect the radio properties in early epoch of tens of days. Such a complex situation will not be explored here.

\subsection{Synchrotron Emission}
The typical frequency of synchrotron radiation from an electron with the LF $\gmm$ is
\be
\nu_m=\frac{\gmm^2 eB}{2\pi m_e c} = 4.08\times 10^7{\rm Hz} ~\gmm^2 \epbt^{\frac{1}{2}} \left(\frac{\dot{m}} {\Omega v_9} \right)^{\frac{1}{2}} t^{-1}_2   
\label{num}
\ee
Approximately, by ${\nu_m} P_{\nu_m} \simeq \frac{4}{3} \sigma_T c \gmm^2 \frac{B^2}{8\pi}$ where $P_{\nu_m}$ is the spectral power at frequency $\nu_m$ from one electron with the LF $\gmm$, we have 
\be
L_{\nu_m} \simeq  N_e \frac{1}{\nu_m} \frac{4}{3} \sigma_T c \gmm^2 \frac{B^2}{8\pi}
\label{lum}
\ee  
where $N_e$ is given by equation \ref{ne}. 
From equation \ref{ne},\ref{bmg},\ref{num},\ref{lum}, $L_{\nu_m}$ is given by
\begin{flalign}
\begin{split}
L_{\nu_m} \simeq \frac{\pi \sigma_T}{3\gmm eB} \kbow \cv  \epb \bar{\epsilon}_{e} \frac{\Mw^2 R_c}{\Omega t^2}  \\
 =5.37\times 10^{31} \ergshz \gmm^{-1} \kbow \cv  \times  \\
\epbt^{\frac{1}{2}} \epeo  \dot{m}^{\frac{3}{2}} \Omega^{-{\frac{1}{2}}} v_9^{{\frac{1}{2}}} R_{c,13} t_2^{-1}  .
\end{split}
\end{flalign}

Synchrotron self-absorption (SSA) absorbs low-frequency radio emission, and determines the peak value of $L_{\nu}$ at the SSA frequency $\nu_a$. The value of $\nu_a$ is derived from $\alpha_{\nu} l$=1, where $\alpha_{\nu}$ is the absorption coefficient and $l$ is the size of emitting region ($\sim R_c$ in our model). In cgs units, 
$\alpha_{\nu} \simeq \frac{\pi^{3/2}}{4} 3^{(p+1)/2} A e B^{-1} \gmm^{-p-4} \nu^{(p+4)/2}_m \nu^{-(p+4)/2}$, where $A$ is given by equation \ref{Acoeff} and $e$ is the electron charge. 
Thus, we have 
\begin{align}
\nu_a &= \nu_m  \left[\frac{\pi^{3/2} 3^{(p+1)/2} (p-1) e \rhow R_c \Lambda } { 4B\gmm^5 m_p}  \right]^{\frac{2}{p+4}}  ~\label{nua0}  \\
=&\begin{cases}
1.63\times 10^8 {\rm Hz} ~f(p) ~ \epbt^{\frac{p+2}{2(p+4)}} \epeo^{\frac{2}{p+4}} \times \\
\left( \frac{\dot{m} }{\Omega v_9} \right)^{\frac{p+6}{2(p+4)}} R^{\frac{2}{p+4}}_{c,13}  ~ t_2^{-\frac{p+6}{p+4}}  ~,~(\vw \leqslant \vdn) \\
3.94\times 10^9{\rm Hz}~ f_2(p)  \epbt^{\frac{p+2}{2(p+4)}} \epeo^{\frac{2(p-1)}{p+4}} \times \\
\left(\frac{\dot{m}}{\Omega} \right)^{\frac{p+6}{2(p+4)}} \beta^{\frac{7p-22}{2(p+4)}} R_{c,15}^{\frac{2}{p+4}} t_2^{-\frac{p+6}{p+4}}  ,~(\vw > \vdn) \label{nua} \\
\end{cases}
\end{align}
where $f(p) =\left[2.6\times 10^{16} \times 3^p (p-1)^2 \right]^{\frac{1}{p+4}}$ and $f_2(p)=\left[ 160 \times 3^p (p-1)^2\right]^{\frac{1}{p+4}}$ for cgs units. As $p$ goes from 2.1 to 3.6 which covers most cases, $f(p)$ decreases monotonically from 739 to 313, while $f_2(p)$ increases monotonically from 3.46 to 4.22. 
For $\nu_a > \nu_m$, the peak synchrotron luminosity at $\nu_a$ can be given by $L_{\nu_a}=L_{\nu_m} (\nu_a/\nu_m)^{(1-p)/2}$. \\
1) When $v \leqslant \vdn$ (deep-Newtonian regime), 
\be
\begin{split}
L_{\nu_a} 
=2.7\times 10^{31} ~\ergshz  \cdot  f(p)^{\frac{1-p}{2}}  ~ \kbow \cv  \times  ~~ \\
\epbt^{\frac{2p+3}{2(p+4)}} \epeo^{\frac{5}{p+4}} \dot{m}^{\frac{2p+13}{2(p+4)}} 
\Omega^{-\frac{5}{2(p+4)}} v^{\frac{2p+3}{2(p+4)}}_9 R_{c,13}^{\frac{5}{p+4}} t_2^{-\frac{5}{p+4}}  ~~.
\end{split}
\label{Lnua}
\ee
2) When $v > \vdn$, 
\be
\begin{split}
L_{\nu_a} =1.28\times 10^{31} ~\ergshz  \cdot  f_2(p)^{\frac{1-p}{2}} ~ \kbow \cv  \times  ~~ \\
 \epbt^{\frac{2p+3}{2(p+4)}} \epeo^{\frac{5(p-1)}{p+4}} \dot{m}^{\frac{2p+13}{2(p+4)}} 
\Omega^{-\frac{5}{2(p+4)}} \beta^{\frac{22p-37}{2(p+4)}} R_{c,13}^{\frac{5}{p+4}} t_2^{-\frac{5}{p+4}} ~~.
\end{split}
\label{Lnua2}
\ee 

Considering SSA, the synchrotron spectrum above $\nu_m$ can be approximately written as
\begin{align}
L_{\nu}=\begin{cases}
L_{\nu_a} (\nu/\nu_a)^{5/2}     ~~~~~~~~~~~~~~ (\nu < \nu_{a}) ~~~ \\
L_{\nu_a} ({\nu/\nu_a})^{(1-p)/2}  ~~~~~~~~~ ( \nu \geqslant \nu_{a})
\end{cases}
\end{align}

\section{Applications}
\subsection{Simplification of the Formula} 
The expressions of $\nu_a$ and $L_{\nu_a}$ depend on several parameters, including $\dot{m}$, $v_9$ (or $\beta$), $\Omega$, $R_{c}$ and $\cv$. We can reduce the parameters under two assumptions.  
First, we assume that $R_c =\eta r= \eta\vw t$, where $\eta$ is a constant. Considering that the typical BLR cloud size is $\sim 10^{13-14}$ cm (\citealt{netzer2015}), and the distance to the BH is $\sim 10^{0-1}$ light days, $\eta$ may be in the order of $10^{-3}- 10^{-2}$. Thus, $\eta=10^{-3}$ is adopted as the fiducial value.  
Second, we naively treat $\Omega \sim 4$ sr as the fiducial value, which corresponding to a biconical outflow with half-opening angle of $47^{\circ}$ (e.g., \citealt{curd2019}). The solid angle is then written as $\Omega \equiv \Omega_4 4{\rm sr}$.
Equations \ref{nua} -- \ref{Lnua2} in this situation can be simplified for $2.1 \leqslant p \leqslant 3.6$. 
\\1) For $v \leqslant \vdn$ (deep-Newtonian regime):
\be
\begin{split}
\nu_a=8.2\times 10^7{\rm Hz} ~ f(p)~ 0.432^{\frac{2}{p+4}} ~\etat^{\frac{2}{p+4}}  \epbt^{\frac{p+2}{2(p+4)}} \epeo^{\frac{2}{p+4}} \times ~~~~~~~~~ \\
 t^{-1}_2 \left( \frac{\dot{m}}{v_9 \Omega_4} \right)^{\frac{1}{2}} \left( \frac{\dot{m} v_9}{\Omega_4} \right)^{\frac{1}{p+4}} ~~~~~~~~~~~~~~~~~~~~ \\
\simeq 3.6\times 10^{10}{\rm Hz} ~ \left( \frac{p}{2.5} \right)^{-1.5}  \eta^{\frac{2}{p+4}}_{-3}  \epbt^{\frac{p+2}{2(p+4)}} \epeo^{\frac{2}{p+4}}   \left( \frac{\dot{m}}{v_9 \Omega_4} \right)^{\frac{1}{2}}   t^{-1}_2   ~,
\end{split}
\label{smnua}
\ee
\be
\begin{split}
L_{\nu_a}=2.7\times 10^{31}{\rm erg ~s^{-1}~Hz^{-1}} ~ f(p)^{\frac{1-p}{2}} 0.432^{\frac{5}{p+4}} \times ~~~~~~  \\
 \kbow \cv \etat^{\frac{5}{p+4}}  \epbt^{\frac{2p+3}{2p+8}} \epeo^{\frac{5}{p+4}} 
 (\dot{m} v_9)^{\frac{2p+13}{2p+8}} \Omega^{-\frac{5}{2p+8}}_4   ~~~~~~~~~~~ \\ 
~~~~ \simeq 1.3\times 10^{29}~\ergshz  \left[(p-1.5)^{-2.2}-0.135 \right]      \times \\
 \kbow \cv  \eta^{\frac{5}{p+4}}_{-3}  \epbt^{\frac{2p+3}{2p+8}} \epeo^{\frac{5}{p+4}}
 (\dot{m} v_9)^{\frac{2p+13}{2p+8}} \Omega^{-\frac{5}{2p+8}}_4   ~.~~~~~~~~~~
\end{split}
\label{smLnua}
\ee
\\2) For $v > \vdn$:
\be
\begin{split}
\nu_a = 1.97 \times 10^9{\rm Hz}~ f_2(p) 168^{\frac{1}{p+4}} \etat^{\frac{2}{p+4}} \epbt^{\frac{p+2}{2(p+4)}} \epeo^{\frac{2(p-1)}{p+4}} \times ~~~~~~~~~~~\\
\left(\frac{\dot{m}}{\Omega_4} \right)^{\frac{p+6}{2(p+4)}} \beta^{\frac{7p-18}{2(p+4)}} t_2^{-1} ~~~~~~~~~~~~~~~ \\
\simeq 1.64 \times 10^{10}{\rm Hz} ~\etat^{\frac{2}{p+4}} \epbt^{\frac{p+2}{2(p+4)}} \epeo^{\frac{2(p-1)}{p+4}} \left(\frac{\dot{m}}{\Omega_4} \right)^{\frac{p+6}{2(p+4)}} \beta^{\frac{7p-18}{2(p+4)}} t_2^{-1}  ~,
\end{split}
\label{smnua2}
\ee
\be
\begin{split}
L_{\nu_a} =1.28\times 10^{31} ~\ergshz  \cdot  f_2(p)^{\frac{1-p}{2}} 13^{\frac{5}{p+4}} \times ~~~~~ \\
~~ \kbow \cv \etat^{\frac{5}{p+4}}  \epbt^{\frac{2p+3}{2(p+4)}} \epeo^{\frac{5(p-1)}{p+4}} \dot{m}^{\frac{2p+13}{2(p+4)}}  
\Omega_4^{-\frac{5}{2(p+4)}} \beta^{\frac{22p-27}{2(p+4)}} ~~~~ \\
\simeq 3.4\times 10^{31}~\ergshz ~ \left(\frac{p}{2.5} \right)^{-3.1}  \times  ~~~~~~~~~~~~~~~~  \\
\kbow \cv \etat^{\frac{5}{p+4}} \epbt^{\frac{2p+3}{2(p+4)}} \epeo^{\frac{5(p-1)}{p+4}} \dot{m}^{\frac{2p+13}{2(p+4)}} \Omega_4^{-\frac{5}{2(p+4)}} \beta^{\frac{22p-27}{2(p+4)}}   .~~~~
\end{split}
\label{smLnua2}
\ee
In equation \ref{smnua}, the final term $\left( \dot{m} v_9 \Omega^{-1}_4 \right)^{\frac{1}{p+4}}$ is within 0.5--2.0 for $\dot{m}v_9 \Omega^{-1}_4 \in [0.01, 100]$ when $p \gtrsim 2.5$. Thus, we simplify it as unit below. 
In deriving the above simplified expressions, we adopt the power-law approximates: $0.432^{\frac{2}{p+4}}f(p)\simeq 4.43\times 10^{2}(p/2.5)^{-1.5}$ (relative error $<3\%$ for $2.1 \leqslant p \leqslant 3.6 $), and $0.432^{\frac{5}{p+4}} f(p)^{\frac{1-p}{2}} \simeq 4.9\times 10^{-3}[(p-1.5)^{-2.2}-0.135]$ (relative error $<10\%$), and $f_2(p) 168^{\frac{1}{p+4}}\simeq 8.3$  (relative error $<3\%$), and $f_2(p)^{\frac{1-p}{2}} 13^{\frac{5}{p+4}} \simeq 2.63(p/2.5)^{-3.1}$ (relative error $<9\%$).  

\begin{figure} 
\includegraphics[width=0.98\columnwidth]{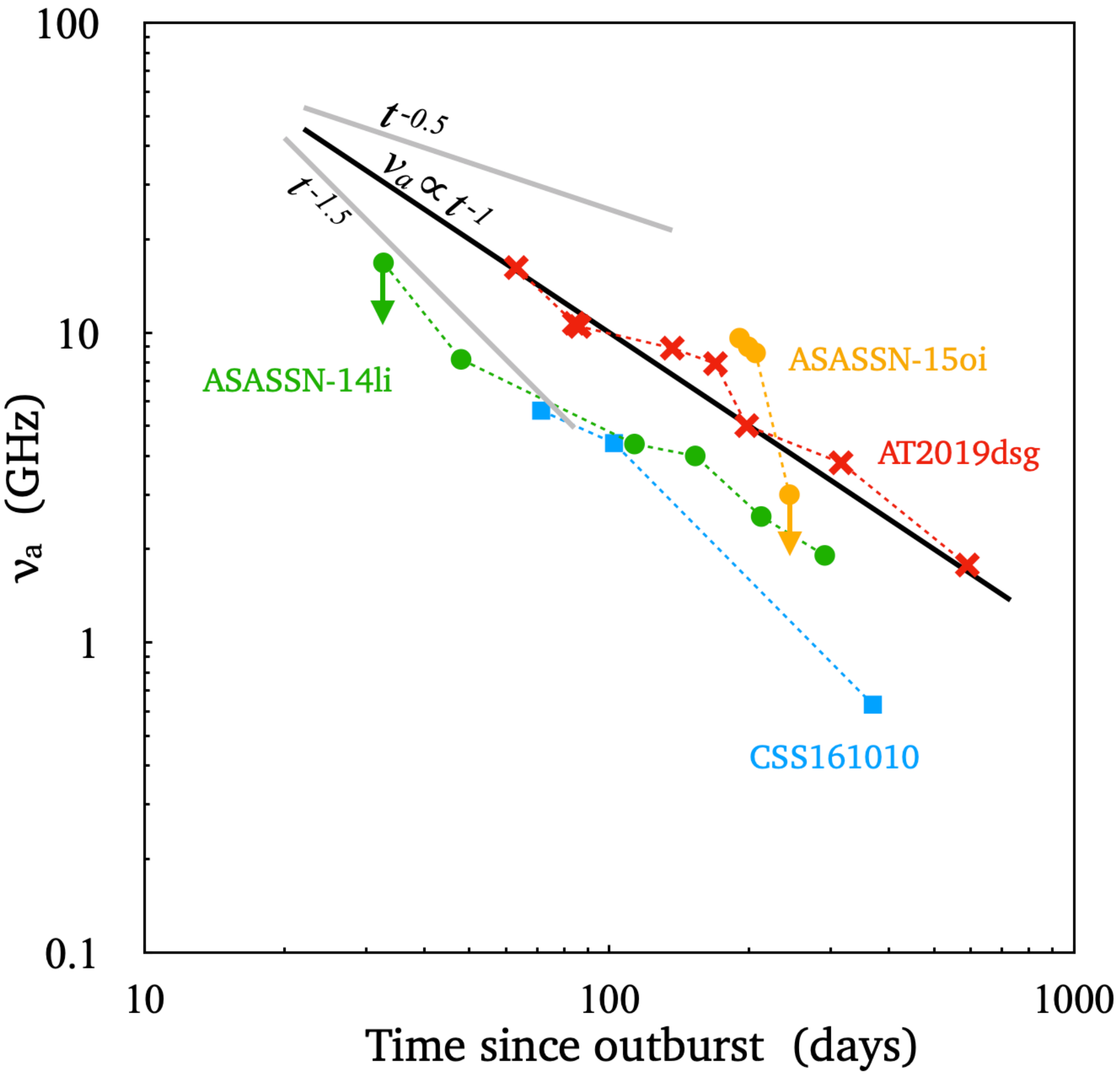}
 \caption{Temporal evolution of the peak frequencies (equivalent to SSA frequencies) of four radio TDEs in the source rest frame: AT2019dsg (\citealt{cendes2021}), ASASSN-14li (\citealt{alexander2016}), CSS161010 (\citealt{coppejans2020}), and ASASSN-15oi (\citealt{horesh2021}). The time intervals are relative to the estimated outburst dates for AT2019dsg and CSS161010 (given in above literatures), while for ASASSN-14li and ASASSN-15oi, the time intervals are relative to the optical discovery dates of 2014-November-22 and 2015-August-14, respectively. We plot the profiles of $\nu_a \propto t^{-1}$, $t^{-0.5}$ and $t^{-1.5}$ in solid lines for comparison. Three of four TDEs roughly follow the law of $\nu_a \propto t^{-1}$, except ASASSN-15oi.  } 
\label{f2va}
\end{figure}

Accordingly, we point out four salient features. 

1. For a TDE outflow of $\dot{m}\sim 1$, $\Omega_4 \sim 1$, and $v_9 \sim 1$, the radio luminosity $\nu_a L_{\nu_a}$ can reach up to $10^{39}~\ergs$, which is enough to account for most radio luminosities (see Figure 1 in \citealt{alexander2020}). When the outflow velocity is higher (e.g., $v_9 \gg1$ or $\beta \sim 1$), or the clouds are more obscuring ($\cv \gg 0.1$) or larger ($\etat \gg1$), $\nu_a L_{\nu_a}$ can reach $\sim 10^{40}~\ergs$ or even higher. 

2. The SSA frequency $\nu_a(t)$ evolves as  $\nu_a \propto t^{-1}$ (a deduction under the assumption of $R_c\propto r$). 
Among the radio TDEs with optical or X-ray outbursts, excluding the jet-induced cases or very late radio flares (arising one year post burst) or those without peak frequencies detections, we obtain four candidates with radio flares arising a few months after outbursts and explicit peak frequencies:  AT2019dsg (\citealt{cendes2021}), ASASSN-14li (\citealt{alexander2016}), CSS161010 (\citealt{coppejans2020}) and ASASSN-15oi (\citealt{horesh2021}). We plot the temporal evolution of the peak frequency (equivalent to $\nu_a$) in Figure \ref{f2va}, and find that three of four sources (except ASASSN-15oi) roughly follow the $t^{-1}$--law. 

3. $L_{\nu_a}$ varies linearly with with $\cv$, implying that the variability of $L_{\nu_a}$ reflects the radial distribution of the cloud's covering factor. When the outflow starts to encounter BLR clouds, $\cv$ increases with time, leading to an increasing $L_{\nu_a}$ over time. 
As the transient outflow expands, the ratio of the outflow ``shell'' thickness to its distance to the BH decreases with $\tbst/t$, and $\cv$ tends to decline over time, resulting in a declining $L_{\nu_a}$. But this is not absolute, due to our poor knowledge of the clouds. 

4. $L_{\nu_a} \propto \Omega^{-\frac{5}{2(p+4)}}$, indicating that when the outflow is more confined, the radio luminosity will be higher. This is in contrast to outflow-CNM model (e.g., \citealt{matsumoto2021}).   

The outflow-BLR model can be extended to outflow-torus scenario, in which the torus is considered to be composed of amounts of dusty clouds. 
In previous works, we initially explored TDE outflow-torus interaction as a possible scenario accounting for years delayed X-ray afterglows (\citealt{mou2021}) and radio afterglows (\citealt{mou2021b}). We found that for an isotropic TDE outflow ($\Omega=4\pi$, called ``TDE wind'' therein) with a kinetic luminosity of $10^{45}~\ergs$, the outflow-torus interaction can yield a considerable radio afterglow. 
Here we can generalize this model in more cases. 
When the outflow kinetic luminosity is of $10^{44}~\ergs$, by setting $c_{f,\rm torus} \sim 0.1$ and $\epbt=\epeo=1$ in equation \ref{smnua}--\ref{smLnua2}, the outflow-torus interaction can generate years delayed ($t_2 \sim 10^1$) radio afterglows with:
1), a peak frequency of $10^9$ Hz which is significantly lower than that of outflow-BLR induced radio flare; 
2), a luminosity of $\nu_p L_{\nu_p} \sim 10^{38}~\ergs$. If including the effects of BLR's blocking and the geometrical light travel, the radio luminosity should  drop by almost one order of magnitude (\citealt{mou2021b}), to $\sim 10^{37}~\ergs$. 
When the adiabatic cooling timescale $t_{ad}=\kbow R_c/\vw \simeq 1{\rm d}~ \etat t_2$ exceeds $\tbst$, the total energy of CRe in equation \ref{ne} will be limited by $t_{\rm bst}$ instead of $t_{\rm ad}$. This is likely to happen for radio emission of years post burst. 
In this situation ($t_{\rm ad}>\tbst$), $L_{\nu_a}$ should be multiplied by a factor of $\tbst/t_{\rm ad}$, and we speculate that $L_{\nu_a}$ decays as $L_{\nu_a} \propto t^{-1}$ (or more rapidly considering a declining $\cv$), while $\nu_a L_{\nu_a}$ decays with $t^{-2}$ (or more rapidly).

\subsection{Constraining TDE Outflow Parameters}

\begin{table*}
\caption{Expressions and Approximations of the Indices ($\ddp\equiv p-2.5$).}
\centering
\setlength{\tabcolsep}{2mm}{
\begin{tabular}{ccccccc} 
\hline \hline 
Index & $\Gamma_1$ & $\Gamma_2$ & $\Gamma_3$ & $\Gamma_4$ & $\Gamma_5$ &$\Gamma_6$  \\

Expression &  $\frac{p+4}{2p+13}$ & $-\frac{9p+46}{(p+4)(2p+13)}$ & $-\frac{2p^2+14p+19}{(p+4)(2p+13)}$ 
& $\frac{p+9}{2p+13}$ & $\frac{3(p+4)}{2p+13}$ & $-\frac{11p+34}{(p+4)(2p+13)}$   \\

Approximation & $0.361+0.014\ddp$ &  $-0.586+0.07\ddp$ & $-0.568-0.05\ddp$ 
& $0.639-0.014 \ddp$ & $1.083+0.04\ddp$ & $-0.526+0.04\ddp$   \\
\hline 

Index & $\Gamma_7$ & $\Gamma_8$ & $\Gamma_9$ & $\Gamma_{10}$  & $\Gm_1$ & $\Gm_2$  \\

Expression & $-\frac{2p^2+8p+5}{(p+4)(2p+13)}$ & $\frac{1-p}{2p+13}$ & $\frac{6-p}{(p+4)(2p+13)}$ & $\frac{3p+7}{(p+4)(2p+13)}$ & $\frac{p+6}{4p+9}$ 
& $-\frac{2p+13}{4p+9}$  \\

Approximation & $-0.321-0.06\ddp$ & $-0.083-0.04\ddp$ & $0.030-0.01\ddp$ & $0.124-0.007\ddp$ & $0.447-0.034\ddp$ 
& $-0.947+0.078\ddp$   \\ 
\hline

Index & $\Gm_3$ & $\Gm_4$ & $\Gm_5$ & $\Gm_6$ & $\Gm_7$ & $\Gm_8$   \\
Expression & $\frac{1}{4p+9}$ & $\frac{-p^2-47p+4}{(p+4)(4p+9)}$ & $\frac{18-7p}{4p+9}$ & $\frac{2(p+4)}{p+6}+\Gm_2 \frac{18-7p}{p+6}$ & $-\Gm_1 \frac{18-7p}{p+6}$ & $-\frac{4}{p+6}-\Gm_3 \frac{18-7p}{p+6}$ \\

Approximation & $0.0526-0.009\ddp$ & $-0.868+0.009\ddp$ & $0.026-0.31\ddp$ & $1.474+0.70\ddp$ & $-0.026+0.31\ddp$ & $-0.474+0.08\ddp$   \\

\hline
Index & $\Gm_9$ & $\Gm_{10}$ & $\Gm_{11}$ \\
Expression & $-\frac{p+2}{p+6}+ \Gm_3 \frac{18-7p}{p+6}$ & $\frac{4(1-p)}{p+6}+\Gm_4\frac{18-7p}{p+6}$ & $1- \Gm_1 \frac{18-7p}{p+6}$  \\
Approximation & $-0.526-0.08\ddp$ & $-0.757+0.29\ddp$ & $0.974+0.31\ddp$ \\

\hline \hline
\end{tabular} }
\label{index}
\end{table*}

For deep-Newtonian regime, from equation \ref{smnua}--\ref{smLnua}, the mass outflow rate, kinetic luminosity and velocity of the outflow are given by (note the symbol $\epeo$, not $\epsilon_{e,-1}$):
\begin{gather}
\dot{m} = \left( \frac{L_{\nu_a}}{\mathcal{L}_1} \right)^{\Gamma_1}  \frac{\nu_a t_2}{\nu_1}   G_1 
(\kbow \cv)^{-\Gamma_1} \etat^{\Gamma_2}  \epbt^{\Gamma_3} \epeo^{\Gamma_2} \Omega_4^{\Gamma_4} , \label{dm1} \\
\dot{m} v^2_9 = \left( \frac{L_{\nu_a}}{\mathcal{L}_1 } \right)^{\Gamma_5} \frac{\nu_1}{\nu_a t_2} G_2 (\kbow \cv)^{-\Gamma_5} \etat^{\Gamma_6} \epbt^{\Gamma_7} \epeo^{\Gamma_6}  \Omega_4^{\Gamma_8} , \\
v_9 = \left( \frac{L_{\nu_a}}{\mathcal{L}_1 } \right)^{\Gamma_1} \frac{\nu_1}{\nu_a t_2} G_3  (\kbow \cv)^{-\Gamma_1}  \etat^{\Gamma_9} \epbt^{\Gamma_{10}} \epeo^{\Gamma_9} \Omega_4^{-\Gamma_1}  , \label{v91}
\end{gather}
where $\mathcal{L}_1=1.3\times 10^{29}~\ergshz$, $\nu_1=3.6\times 10^{10}~{\rm Hz}$, and 
\begin{align}
\begin{cases}
G_1=\left( \frac{p}{2.5} \right)^{1.5} [(p-1.5)^{-2.2}-0.135]^{-\Gamma_1} , \\
G_2=\left( \frac{p}{2.5} \right)^{-1.5} [(p-1.5)^{-2.2}-0.135 ]^{-\Gamma_5} , \\
G_3=\left( \frac{p}{2.5} \right)^{-1.5}[(p-1.5)^{-2.2}-0.135 ]^{-\Gamma_1} . 
\end{cases}
\end{align}
The expressions and approximations of the indices $\Gamma_1 - \Gamma_{10}$ are listed in Table \ref{index}. The approximations ($\ddp\equiv p-2.5$ therein) are suitable for $2.1 \leqslant p \leqslant 3.6$. 

For $\vw > \vdn$,  from equation \ref{smnua2}--\ref{smLnua2}, the velocity and mass outflow rate are given by (again, note the symbol $\epeo$):
\begin{gather}
\beta=\left( \frac{L_{\nu_a}}{\mathcal{L}_2} \right)^{\Gm_1} \left(\frac{\nu_a t_2}{\nu_2} \right)^{\Gm_2}  G_4 ~(\kbow \cv)^{-\Gm_1} \times \nonumber \\
\etat^{-\Gm_3} \epbt^{\Gm_3} \epeo^{\Gm_4} \Omega_4^{-\Gm_1}  ~, \label{beta2} \\
\dot{m}=\left( \frac{L_{\nu_a}}{\mathcal{L}_2 } \right)^{\Gm_5} \left(\frac{\nu_a t_2}{\nu_2 } \right)^{\Gm_6}  G_5  ~
(\kbow \cv)^{\Gm_7} \times \nonumber \\
 \etat^{\Gm_8} \epbt^{\Gm_9} \epeo^{\Gm_{10}} \Omega_4^{\Gm_{11}}  ~, \label{dm2}
\end{gather}
where $\mathcal{L}_2 =3.4\times 10^{31}~\ergshz$, $\nu_2=1.64\times10^{10}~{\rm Hz}$, and
\begin{align}
\begin{cases}
G_4=\left( \frac{p}{2.5} \right)^{3.1\Gm_1} , \\
G_5=\left( \frac{p}{2.5} \right)^{3.1\Gm_5} .
\end{cases}
\end{align}
The indices of $\Gm_1$, $\Gm_2$,  ..., $\Gm_{11}$ are listed in Table \ref{index}, and the approximations are suitable for $2.1 \leqslant p \leqslant 3.6$. Note that our study here concerns the non-relativistic outflow, and in relativistic case, the above calculations will fail. 

$L_{\nu_a}$ can be given by observations via $4\pi D^2_L F_{\nu_{a}}$, where $D_L$ is the luminosity distance, and $F_{\nu_{a}}$ is the observed radio flux at the peak frequency $\nu_a$. The index $p$ can be obtained from radio spectra: $F_{\nu} \propto \nu^{(1-p)/2}$ for $\nu > \nu_a$. 
With observational parameters $\nu_a$, $L_{\nu_a}$ and $p$, and assumptions of $\cv$ and $\eta$, we can obtain rough estimates of the outflow parameters. 
The final mass outflow rate and kinetic luminosity of TDE outflow are transferred by $\Mw=\dot{m} ~\msunyr$, and $\Lkin=3.2\times 10^{43}~\ergs \dot{m}v^2_9$ or $2.9\times 10^{46}~\ergs \dot{m} \beta^2$.

\subsection{Application to TDE Candidates} 

\begin{table*}
\begin{center}
\caption{Radio observations and Constraints on Outflow Physics. 1. $\Delta t$ is counted from the estimated outburst dates (AT2019dsg, CSS161010) or optical discovery date (ASASSN-14li) in the observer frame.  2. AT2019dsg: $z=0.051$, $D_L=230$Mpc; refer \citet{cendes2021} for radio data.  3. ASASSN-14li: $z=0.0206$, $D_L=90$Mpc; refer \citet{alexander2016} for radio data. 
 4. CSS161010: $z=0.034$, $D_L=150$Mpc; refer \citet{coppejans2020} for radio data. } 
\setlength{\tabcolsep}{2mm}{
\begin{tabular}{ccccc  ccc  ccc  ccc}
\hline \hline
\multicolumn{5}{c}{  } &\multicolumn{3}{c}{($\Omega_4=1$, ~~~$\etat=1$)} &\multicolumn {3}{c}{$(\Omega_4=3.14$, ~~~$\etat=1$)} &\multicolumn {3}{c}{$(\Omega_4=1$, ~~~$\etat=10)$} \\
Name  & $\Delta t$ & $\nu_p$ & $F_{\nu_p}$ & $p$ & $\Mw$   & $\Lkin$    & $\vw$ & $\Mw$   & $\Lkin$    & $\vw$ & $\Mw$   & $\Lkin$    & $\vw$ \\
           &  (d)       &  (GHz)   &   (mJy)         &        & ($\msunyr$) & ($\ergs$) &  ($c$)   & ($\msunyr$) & ($\ergs$) &  ($c$)   & ($\msunyr$) & ($\ergs$) &  ($c$)    \\
\hline
AT2019dsg  & 60  &  16.2  & 0.74 & 2.7& 0.28 & $6.6\times 10^{43}$ & 0.09  &  0.57 & $5.9\times 10^{43}$ & 0.06 & 0.07 & $2.0\times 10^{43}$ & 0.10 \\
ASASSN-14li  & 47  & 8.2 & 1.76 & 3.0 & 0.11 & $1.0\times 10^{44}$ & 0.18 & 0.22 & $8.7\times 10^{43}$ & 0.12 & 0.03 & $3.1\times 10^{43}$ & 0.19 \\
CSS161010    & 69  & 5.6 &  8.8 & 3.6 & 0.14 & $2.0\times 10^{45}$ & 0.70  & 0.62 & $3.5\times 10^{45}$ & 0.44 & 0.06 & $6.8\times 10^{44}$ & 0.64  \\
\hline \hline  
\end{tabular} } 
\end{center}
\label{tab2}
\end{table*} 

We apply the model to three candidates: AT2019dsg, ASASSN-14li\footnote{For ASASSN-14li, AGN-like emission line spectrum has been detected in the off-nuclear region, suggesting a recent shut-off of nuclear activity (\citealt{prieto2016}).} and CSS161010. 
To estimate the outflow parameters, we select their earliest radio data that fully covering the peak frequencies (see Table \ref{tab2}): 2019-May-29 for AT2019dsg (\citealt{cendes2021}), 2015-Jan-6 and Jan-13 for ASASSN-14li (\citealt{alexander2016}), and 2016-Dec-14 for CSS161010 (\citealt{coppejans2020}).  
By assuming $\epbt = \epeo = \kbow \cv = 1$, the mass outflow rates, kinetic luminosities and velocities of TDE outflow can be estimated via equation \ref{dm1}--\ref{v91} (AT2019dsg and ASASSN-14li), or \ref{beta2}--\ref{dm2} (CSS161010), which are listed in Table \ref{tab2}. Here we explored three cases of ($\Omega_4, \eta_{-3})=$ (1,1), (3.14, 1) and (1,10). 

The estimated outflow velocities for these sources are indeed ultra-fast, ranging from 0.06c to 0.7c. 
Despite the great uncertainties (due to several unknown parameters), the inferred outflow kinetic luminosities of $10^{43-45}~\ergs$ are consistent with those in GRRMHD simulations ($\sim 10^{44-45}~\ergs$, \citealt{curd2019}). 
Accordingly, we argue that if the fast and energetic TDE outflow ($\gtrsim 10^{44}~\ergs$) predicted by theoretical studies is prevalent, whether it yields a radio flare depends on the existence of cloudy circumnuclear environment. 
In other words, the detection rate of radio flares in TDE (\citealt{alexander2020}) implies that the proportion of cloudy circumnuclear environment (covering factor reaching the order of 0.1 on the BLR scale) is a few percents. 
If the outflow continues to emanate for $1\sim2$ months, the ejecta mass $M_{\rm out}$ is $10^{-2}$ to $10^{-1}$ solar mass, and the total kinetic energy is in the order of $10^{50-52}$ erg. This suggests that a considerable fraction of infalling materials forms fast outflow, considering the median mass of $\sim 0.3\msun$ in the initial mass function (\citealt{kroupa2001}). 
Moreover, our kinetic energy is also consistent with the predictions of collision-induced outflow scenario (\citealt{lu2020}). 

For AT2019dsg, outflow-CNM model suggests that the outflow velocity is 0.12c (\citealt{stein2021}) or 0.07c (\citealt{cendes2021}),
which is coincidentally close to our results.  
We note a recent report on this source exhibiting strong infrared echo arising almost simultaneously with optical burst (\citealt{van2021}), suggesting that clouds exist near the central BH. 
ASASSN-14li was detected with an ionized outflow with a velocity of 0.2c within 40 d after discovery (\citealt{kara2018}), which is consistent with our result. In contrast, outflow-CNM model reports a velocity of 0.04--0.12c, and a kinetic energy of $4-10\times 10^{47}$ erg  (\citealt{alexander2016}), which are significantly lower than ours. 
CSS161010 was estimated to launch an outflow with initial velocity $\Gamma \beta c \geqslant 0.55c$, kinetic energy $\gtrsim 10^{51}$ erg and ejecta mass $0.01-0.1\msun$ by outflow-CNM model, which are also coincidentally similar to our results. 
It's worth noting that, CSS161010 is a fast blue optical transient residing in a dwarf galaxy of 10$^7\msun$ (\citealt{coppejans2020}), and the central BH should be very small if it does exist. Thus, our kinetic luminosity is much higher than its Eddington luminosity (also true for optical luminosity and outflow-CNM model deduced outflow, \citealt{coppejans2020}), which makes this object very puzzling.  


\section{Discussions} 
\begin{figure}
\includegraphics[width=0.9\columnwidth]{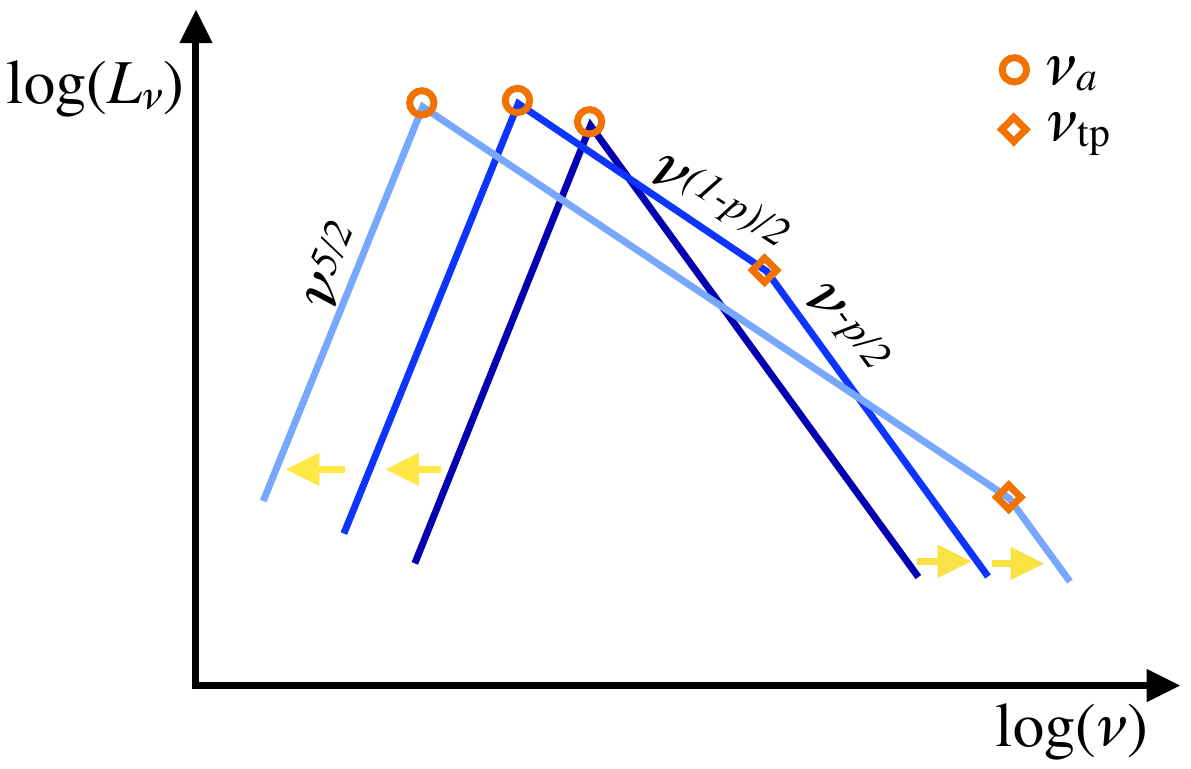}
 \caption{ A sketch of the radio spectrum evolution when incorporating synchrotron cooling. The yellow arrows mark the direction of evolution (from dark blue to blue, to light blue). The dark blue case ($\nutp<\nu_a$) only appears under some conditions, e.g., large clouds ($\etat \gtrsim 10$) and early stage ($t_2 \sim 10^{-1}$). }  
 \label{fig3}
\end{figure}

Due to the strong magnetic field at bow shocks (equation \ref{bmg}), spectrum of CRe with LF above $\gamma_{\rm tp}$ is one power of $\gamma_e$ steeper ($dN(\gamma_e)/d\gamma_e \propto \gamma^{-p-1}_e$), where the turning point $\gamma_{\rm tp}$ is determined by synchrotron cooling timescale. Since accelerated electrons at bow shock cannot survive after the adiabatic cooling timescale of $t_{\rm ad}\simeq \kbow \Rc/\vw=8.6\times 10^4 \etat t_2$, synchrotron cooling only affects those electrons that have not yet adiabatically cooled. By equating $t_{\rm ad}$ and $t_{\rm syn}$ ($t_{\rm syn}=8\times 10^8~{\rm s}~  \gamma^{-1} B^{-1} =3.8\times 10^{6}~{\rm s}~ \gamma^{-1} \epbt^{-1} \dot{m}^{-1} v_9 t_2^2$), we have $\gamma_{\rm tp} = 1.8\times 10^2 \epbt^{-1}\etat^{-1} \dot{m} ^{-1} \Omega_4 v_9 t_2 $. Accordingly, the turning point frequency for $\gamma_{\rm tp}$ is 
\be
\nutp=6.6\times 10^{11} {\rm Hz}~ \etat^{-2} \epbt^{-\frac{3}{2}} \left(\frac{\dot{m}}{v_9 \Omega_4} \right)^{-\frac{3}{2}} t_2  ~, 
\ee
which increases over time (in contrary to $\nu_a$), and is sensitive to the cloud size parameter $\etat$. 
When $\nutp >\nu_a$, the radio spectrum above $\nu_a$ will present two sections (see Figure \ref{fig3}): slope of $(1-p)/2$ below $\nutp$, and slope of $-p/2$ above $\nutp$. 
When $\nutp < \nu_a$, the slope of radio spectrum above $\nu_a$ becomes $-p/2$ instead of $(1-p)/2$. 
The condition of $\nutp < \nu_a$ is $\etat t^{-1}_2 \epbt \dot{m} \Omega^{-1}_4 v^{-1}_9  \gtrsim 4.3 (p/2.5)^{0.75}$ for $v\leqslant v_{\rm DN}$, or $\etat t^{-1}_{2} \epbt \epeo^{(p-1)/(p+4)} \dot{m} \Omega_4^{-1} \beta^{(2p-15)/(2p+8)} \gtrsim 81$ for $v>v_{\rm DN}$ (approximation for $2.1\leqslant p \leqslant 3.6$), which can be satisfied for large clouds ($\etat \gtrsim 10$) and radio flares in early stage ($t_2 \sim 10^{-1}$). 
Qualitatively (not strictly, as $\etat$ may vary with distance $r$), cooling break in radio spectrum tends to appear in early stage and disappear in late stage as $\nutp$ growing beyond the observed frequency. This is contrary to outflow-CNM model, in which $\nutp$ is expected to decrease over time due to synchrotron cooling. 

Which process dominates the overall radio emission when clouds coexist with CNM?
It depends on two factors: the shock energy (or CRe energy) and the magnetic field.
The density of CNM around Sgr A* follows $300~ \cmc r^{-1}_{-2}$ ($r_{-2} \equiv r/10^{-2}$pc, \citealt{xu2006,gillessen2019}). 
Here we simply assume a CNM density following $n(r)=\Csgr \times 300 ~\cmc r^{-1}_{-2}$ or $n(r)=\Csgr \times 300 ~\cmc r^{-2}_{-2}$, where $\Csgr$ is a constant. 
The enclosing CNM mass within $r$ and the solid angle $\Omega$ is $\mcnm(r)=\int n(r) m_{\rm H} \Omega r^2 dr=1.5\times 10^{-5}\Csgr   \Omega_4 r^2_{-2} \msun$ ($n\propto r^{-1}$) or $3.0\times 10^{-5} \Csgr \Omega_4 r_{-2} \msun$ ($n\propto r^{-2}$). 
Since the post-shock CNM roughly co-moves with TDE outflow, the energy of forward shock is $E_{\rm forwd}\sim \Lkin t \cdot \mcnm(r)/M_{\rm out}$ when $\mcnm(r)<M_{\rm out}$. 
On the other hand, limited by adiabatic cooling, the bow shock energy is $E_{\rm bow}\sim \Lkin \cv \kbow R_c/\vw = \Lkin \cv \kbow \eta t$. Thus, the ratio of the two shock energies is
\begin{align}
\frac{E_{\rm bow}}{E_{\rm forwd}}  =\begin{cases}
\frac{6.7 \kbow \cv \etat}{\Csgr \Omega_4 r_{-2}^2} \cdot \frac{M_{\rm out}}{0.1\msun} ~~ (n\propto r^{-1})  \\
\frac{3.3 \kbow \cv \etat}{\Csgr \Omega_4 r_{-2}}    \cdot \frac{M_{\rm out}}{0.1\msun} ~~ (n\propto r^{-2})
\end{cases}
\end{align} 
which relies on several parameters including CNM and the outflow. 
Roughly speaking, for a dense CNM ($\Csgr >10$) and less cloudy environment ($\kbow \cv  \etat \lesssim 1$), the total energy of CRe would be dominated by forward shock in outflow-CNM interaction. 

The magnetic field strengths in two scenarios are quite different. At the forward shock, $B\simeq (\epb 8\pi m_p n \vw^2)^{0.5}=0.021 {\rm Gauss}~ \epbt^{0.5} \Csgr^{0.5} v_9^{0.5} t_2^{-0.5}$ for $n\propto r^{-1}$, and $0.04~{\rm Gauss}~ \epbt^{0.5} \Csgr^{0.5} t_2^{-1}$ for $n\propto r^{-2}$. 
Thus, the magnetic field ratio of bow and forward shock is (assuming that values of $\epbt$ are equivalent for both shocks) 
\begin{align}
\frac{B_{\rm bow}}{B_{\rm forwd}} =\begin{cases}
348 \Csgr^{-0.5}\dot{m}^{0.5} \Omega_4^{-0.5} v_9^{-1} t_2^{-0.5} (n\propto r^{-1}) \\
182 \Csgr^{-0.5} \dot{m}^{0.5} \Omega_4^{-0.5} v_9^{-0.5} ~~~~(n\propto r^{-2})
\end{cases}
\end{align} 
In a wide parameter space, the magnetic field of the bow shock is stronger than that of the forward shock. 
Since $L_{\nu}$ is proportional to the magnetic field strength, even if the CRe energy of the forward shock mildly exceeds that of the bow shock, the overall radio emission may still be dominated by the bow shock. 
We can calculate a few examples to illustrate this point. 

\begin{figure}
\includegraphics[width=0.98\columnwidth]{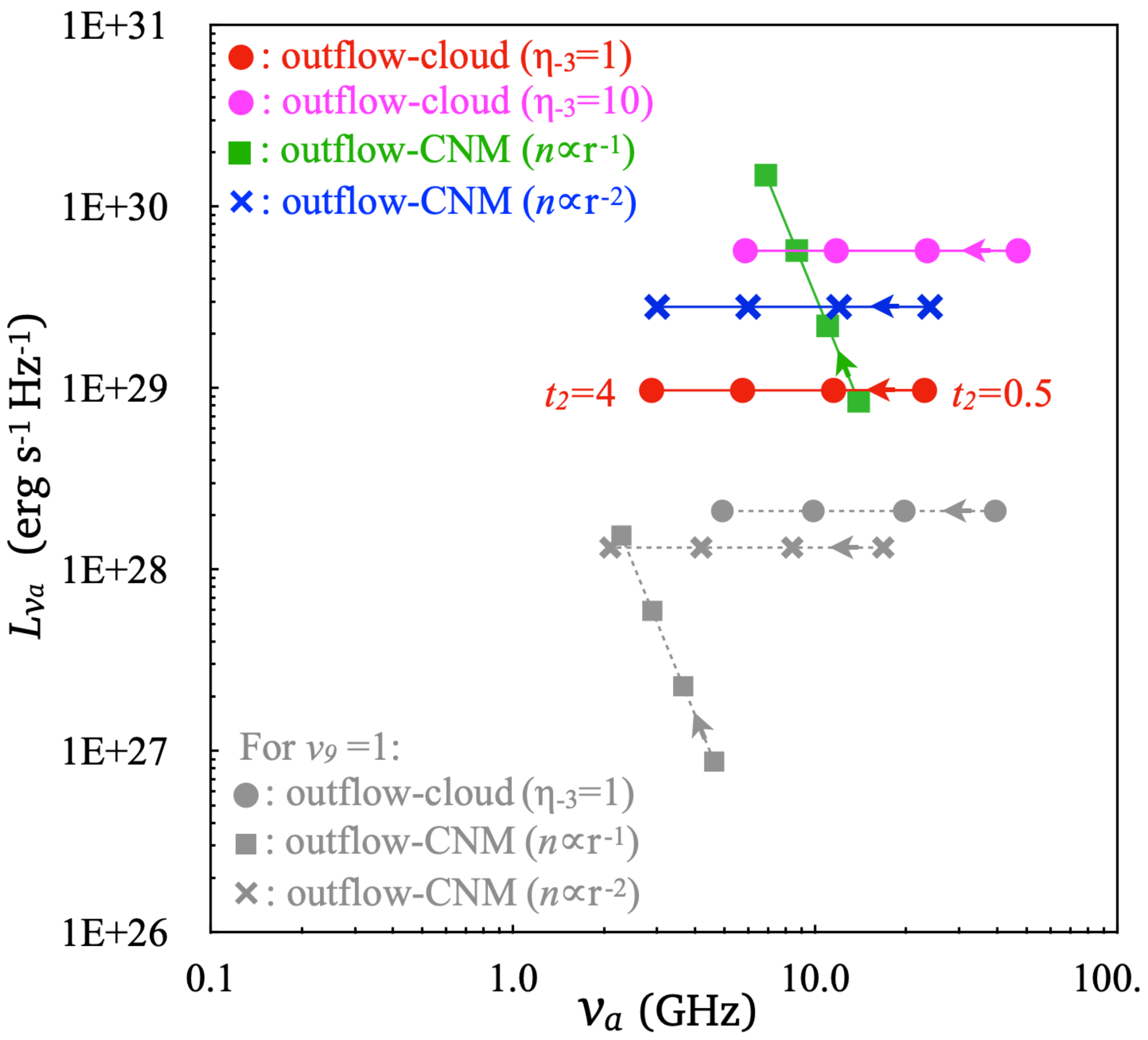}
 \caption{Temporal evolution of $\nu_a$ and $L_{\nu_a}$ for outflow-cloud and outflow-CNM interaction
with $\Csgr=100$. 
We fixed the following parameters for all cases: $\Omega_4=\epbt=\epeo=1$ and $p=2.5$. 
The arrows mark the direction of evolution ($t_2$ from 0.5 to 1, 2, and 4). 
The circles represent the outflow-cloud process with $\dot{m}=0.3$ and $\kbow \cv=1$ (red for $\etat=1$, magenta for $\etat=10$). 
The squares and crosses denote the outflow-CNM interaction with densities following $r^{-1}$ and $r^{-2}$, respectively. 
The colored and grey symbols mark the outflow velocity of $v_9=3$ and 1, respectively. Even for a dense CNM of $\Csgr=100$, the outflow-cloud induced radio emission can be comparable to the outflow-CNM induced one, or even exceeds the latter if the clouds are large ($\etat \sim 10$) or the outflow is slow ($v_9<3$). }  
 \label{fig4}
\end{figure}

Following the calculations in \citet{matsumoto2021}, we draw the temporal evolutions of the peak frequencies and luminosities for outflow-CNM process in Figure \ref{fig4} with $\Csgr=100$ and $p=2.5$. To obtain the evolution curves of outflow-CNM model 
for other $\Csgr - $values, one can simply translate the curves by $\nu_a \propto \Csgr^{0.65}$ and $L_{\nu_a} \propto \Csgr^{1.4}$. 
In a mildly dense CNM environment of $\Csgr\lesssim 100$, radio from outflow-cloud interaction may be non-negligible, or even dominate the overall radio emission. 
Furthermore, when the outflow velocity is relatively low ($\vw < 0.1$c), the radio emission favors the outflow-cloud interaction in more cases. 
Through these examples, we speculate that, there is a ``critical'' density ratio 
roughly in the order of $\sim 100$, below which outflow-cloud interaction may become non-negligible in radio emission. 
Moreover, for $n\propto r^{-1}$, radio emission in different epochs may come from different processes: outflow-cloud in early stage and outflow-CNM in late stage. 

\section{Conclusions} 
TDE outflow driven into the ambient environment will generate shocks and give rise to radio emission when encountering the diffuse CNM or condensed clouds. We explored the outflow-cloud interaction in active galaxies, which is manifested as outflow-BLR interaction, and can be extended to outflow-torus interaction. Such an interaction is able to generate considerable radio emission, of which the radio luminosity can reach up to $10^{39}~\ergs$ or even higher (equation \ref{smnua}--\ref{smLnua2}). 
Moreover, this process predicts that the SSA frequency declines as $\nu_a \propto t^{-1}$, which is observed in some radio TDEs \footnote{Such a decline law can also appear in outflow-CNM process if the CNM density satisfies $n(r)\propto r^{-2}$ (e.g., \citealt{matsumoto2021}). }.
Eventually (a few years post burst), the luminosity $\nu_a L_{\nu_a}$ decays in a law of $t^{-2}$ or more steeper. 
The radio emission relies on the bow shock energy, which reflects the properties of the outflow and clouds. Thereby, it provides a method for directly constraining the physics of the outflow (equation \ref{dm1}--\ref{v91}, \ref{beta2}--\ref{dm2}), although very rough due to several unknown parameters. For the outflow-CNM interaction, however, the radio reveals the properties of CNM, which is adopted to indirectly constrain the outflow's physics. 
The outflow-cloud model is applied to three radio TDEs (including the puzzling CSS161010),
and the inferred outflow kinetic luminosities of $10^{43-45} ~\ergs$ are consistent with simulations on the settled super-Eddington accretion disk (\citealt{curd2019}). 
When incorporating synchrotron cooling, cooling break in radio spectrum tends to appear in early stage and disappear in late stage.   

By comparing the radio intensity in both outflow-cloud and outflow-CNM processes, we find that, 
in a mildly or less dense CNM ($\lesssim 100 \times$ Sgr A*-like CNM) and a cloudy environment, the contribution of outflow-cloud induced radio may be non-negligible or even dominant in the overall radio emission, especially for relatively large clouds ($\etat > 1$), or relatively low outflow velocities ($\vw \lesssim$ 0.1c). Moreover, both outflow-cloud and outflow-CNM induced radio emission may coexist in one source, but manifest separately in the early and late epochs. The increasing radio samples for TDEs will help us examine the mechanisms for radio flares. If it is prevalent in radio flares, outflow-cloud induced radio may be also used to study whether ``hidden'' clouds (due to the non-illuminated before TDE) exist in the vicinity of quiescent SMBHs.


\section*{Acknowledgements}
We are grateful to the anonymous referee for providing comments that significantly improved the manuscript. G.M. is supported by National Science Foundation of China (11833007 and 11703022). 
T.W. is supported by NSFC through grant NSFC-11833007 and 11421303.
W.W. is supported by the NSFC under grant number 12133007, U1838103.

\section*{Data Availability}
The data underlying this article will be shared on reasonable request to the corresponding author. 
{\bf A program for calculating outflow parameters of this model is available on github 
(https://github.com/G-Mou/RadioTDE) or CSDN (https://blog.csdn.net/Swift\_csdn/article/details/121706448). }




\end{document}